\documentclass[a4paper]{article}
\usepackage{graphicx}
\usepackage{chicago}
\newcommand{\be}{\begin{equation}}
\newcommand{\ee}{\end{equation}}
\newcommand{\bd}{\begin{displaymath}}
\newcommand{\ed}{\end{displaymath}}
\newcommand{\bea}{\begin{eqnarray}}
\newcommand{\eea}{\end{eqnarray}}
\newcommand{\bi}{\begin{description}}
\newcommand{\ei}{\end{description}}
\newcommand{\bq}{\begin{quote}}
\newcommand{\eq}{\end{quote}}
\def\i{\item}
\def\fo{\footnote}

\def\ni{\noindent}
\def\ua{\uparrow}
\def\da{\downarrow}
\def\ra{\rightarrow}

\def\r{\rho}
\def\a{\alpha}

\def\e{\epsilon}

\def\om{\omega}
\def\Om{\Omega}

\def\S{\Sigma}

\def\ph{\varphi}
\begin{document}
\bibliographystyle{chicago}
\twocolumn[
\author{Alexander~Unzicker\\
        Pestalozzi-Gymnasium  M\"unchen\\[0.6ex]
{\small{\bf e-mail:}  alexander.unzicker@lrz.uni-muenchen.de}}
\date{October 6, 2007}
\title{Mach's Principle and a Variable Speed of Light}
\maketitle

\begin{abstract}
Ernst Mach (1838-1916) suggested that the origin of gravitational interaction
could depend on the presence of all masses in the universe.
A corresponding hypothesis of Sciama (1953) on the gravitational constant,
$c^2/G = \sum m_i/r_i$, is linked to Dicke's (1957)
proposal of an electromagnetic origin of gravitation, a
precursor of scalar-tensor-theories. In this an equivalent description
in terms of a variable speed of light (VSL) is given, and the
agreement with the four classical tests of general relativity is shown.
 Moreover, VSL  opens the
possibility to write the total energy of a particle as $E=mc^2$;
this necessarily leads to the proportionality of inertial and gravitating
mass, the equivalence principle.
Furthermore, a formula for $c$ depending on the mass distribution is given
that reproduces Newton's law of gravitation.\footnotemark 
This mass distribution allows to calculate
a slightly variable term that corresponds to the `constant' $G$.
The present proposal may also supply an alternative
explanation to the flatness problem and the horizon problem in cosmology.

\end{abstract}
\vspace{1.0cm}]

\footnotetext{I do not uphold any longer the 
formula given here, and prefer the proposal outlined in arxiv:0708.3518 \citeyear{Unz:07c} 
which additionally matches Dirac's large number hypothesis \citeyear{Dir:38a}.} 

\section{Introduction}
\subsection{Overview}
After a brief motivation given in 1.2 I shall outline in sec.~2
 how Mach's principle can be realized
as a quantitative statement, following \citeN{Sci:53}.
Before Sciama's theory is related to a
variable speed of light, in sec.~3 general
considerations on time and length scales and a brief historical perspective
of VSL theories are given.
A very similar approach of describing gravity as an electromagnetic
effect has been outlined in Dicke's \citeyear{Dic:57} paper, a precursor of the
so-called scalar-tensor theory.
Using similar
arguments as \citeN{Wil}, sec.~4 repeats
how the respective experimental tests of general relativity (GR)
can alternatively be described by a variable $c$. While this is
merely another viewpoint
of classical physics,  Sec.~5 describes as a new proposal how
Newton's law of gravitation can be directly deduced from a variable $c$.
While this general approach has interesting consequences
for the equivalence principle, in sec.~6 an explicit formula for $c$ depending on mass
distributions is given  that is in accordance with Mach's principle.
In sec.~7, consequences are discussed with respect to observational data.

\subsection{Open questions in gravitational physics}

While the theory of general relativity (GR), at the time of its discovery,
could predict the last observation incompatible with Newtonian gravity,
the mercury perihelion,
new riddles
have shown up in gravitational physics in the past decades.
Since Zwicky's observation
of `missing mass' in galaxy clusters an overwhelming evidence for
`dark matter' has been collected, in particular the flat rotation curves of
galaxies. Since 1998, the relatively faint high-redshift
supernovae are commonly explained as a manifestation of a new  
form of matter called `dark energy'. As \shortciteN{Agu:01a} comment, these new
discoveries have been achieved "at the expense of simplicity"; it
is quite disappointing that no counterpart to these forms of matter
has been found in the laboratories yet. Thus in the view of recent data
the following comment of \citeN{BinGD}, p.~635, is more than justified:
\bq
`It is worth remembering
that all of the discussion so far has been based on the premise that
Newtonian gravity and general relativity are correct on large scales.
In fact, there is little or no direct evidence that conventional
theories of gravity are correct on scales much larger than a parsec
or so. Newtonian gravity works extremely well on scales of
$\sim 10^{14} cm$ (the solar system). (...)  It is principally the
elegance of general relativity and its success in solar system
tests that lead us to the bold extrapolation that the gravitational
interaction has the form $GM/r^2$ on the scales $10^{21}-10^{26} cm$...'
\eq

In the meantime, the bold extrapolation seems to have encountered
new observational problems. The detection of an anomalous
acceleration from the Pioneer missions (\citeNP{And:01};
\shortciteNP{Tur:05}) indicate that Newtonian gravity may not even
be correct at scales within the solar system. Interestingly, this
anomaly occurred at the same {\em dynamic\/} scale - about
$10^{-9} m/s^2$ - as many phenomena indicating dark matter.

Furthermore, the scale $10^{-9} m/s^2$ is the scale that most
of the precision measurements of $G$ approach - it is still discussed
if the  discrepancies between the $G$ measurements in the last
decade (\citeNP{Gun:00}; \citeNP{Uza}) have a systematic reason.

Besides
the distance law, there is another lack of experimental data regarding the mass
dependence. The Cavendish torsion balance uses masses of about $1 kg$
to determine the mass of the earth, since there is no precise independent
geological measure. The two-body treatment of Newtonian gravity tells us that
the relative acceleration $a_{12}$ is proportional to $m_1+m_2$.
It may well be that simple form is just an approximation valid for
$m_1>> m_2$. A slightly different
mass dependence\fo{See the interesting measurements by \shortciteN{Hol:86},
\shortciteN{And:89} and \shortciteN{Zum:91}.}  would remain undetected even by
the most precise emphemeris and double pulsar data, since in most cases just
the product $GM$ is measured.

In summary, there is quite a big gap between the common belief in the universality
of Newton's law of gravitation and the experimental evidence supporting 
it.\fo{A more detailed discussion is given in \citeN{Unz:07}.}
So far there is no evidence at all that it still works for accelerations
below $10^{-9} m/s^2$.

\section{Mach's principle}

Besides the observational facts above, Newtonian gravity had to
face theoretical problems, too. In his famous example of the
rotating bucket filled with water, Newton deduced the existence of
an absolute, nonrotating space from the observation of the curved
surface the water forms. Ernst Mach criticized this `non observable'
concept of absolute space as follows, suggesting that the water
was rotating with respect to masses at large distance: \bq `No one
is competent to say how the experiment would turn out if the sides
of the vessel increased in thickness and mass till they were
ultimately several... [miles]... thick.' \eq Mach's principle is
commonly known as follows: The reason for inertia is that a mass
is accelerated with respect to all other masses in the universe,
and therefore gravitational interaction would be impossible
without the distant masses in the universe.

A possible solution to the `rotating bucket' part of the problem
where the distant masses instead of absolute space define the
physical framework has been given in a brilliant paper by
\citeN{Lyn:95}.

However, from Mach's principle it has been further deduced that
the numerical value of the Gravitational G constant must be
determined by the mass distribution in the universe, while in
Newton's theory it is just an arbitrary constant. Since the square
of the speed of light times the radius of the universe divided by
the mass of the universe is approximately equal to G, such
speculations were first raised by \citeN{Dir:38a}, \citeN{Sci:53}
and \citeN{Dic:57}.

\subsection{Sciama's implementation of Mach's principle.}

Since Mach  never gave a quantitative statement
of his idea, it has been implemented in various manners; the reader
interested in an overview is referred to \citeN{Bon}, \citeN{BaP} and \citeN{Gra}.
An idea of a quantitative statement proposes the following
functional dependence of Newton's constant $G$ (\citeNP{Sci:53}; \citeNP{Unz:03a}):
\be
\frac{c^2}{G} =\sum_{i}  \frac{m_i}{r_i}= \Phi \label{sum}
\ee
As in Sciama's proposal, (\ref{sum}) may be
included in the definition of the gravitational potential,
thus yielding
\be
\ph= -G \, \sum_{i} \frac{m_i}{r_i} = -G \Phi = -c^2. \label{pot}
\ee

Alternatively, one can apply the hypothesis (\ref{sum}) directly to Newton's
law. The potential
instead can then be brought to the form
\be
\varphi = -c^2 \ln \sum_{i} \frac{m_i}{r_i}   \label{logpot}
\ee
from which follows Newton's law
\be
\label{sci}
-\nabla \varphi = -c^2 \frac{\sum_{i}
\frac{m_i \vec r_i}{r_i^3}}{\sum_{i} \frac{m_i}{r_i}} =: -G(m_i, r_i) \sum_{i} \frac{m_i \vec r_i}{r_i^3}.
\ee

This was already developed by \citeN{Unz:03a}, though I do not
support any more the further considerations following  there. I
prefer instead, as (\ref{pot}) suggests, to relate the
(necessarily varying) gravitational potential to a variable $c$.
It seems that Sciama did not consider this possibility; an
explicit form will be given in sec.~\ref{sciama2} which has
however similarities with (\ref{logpot}). For systematic reasons
however, we should analyze the consequences of a variable $c$ at
first.

\section{A spacetime with variable $c$.}

\subsection{Early attempts of Einstein.}
Before further relating $c$ to the gravitational potential
 it is worth remembering that a possible influence
of the speed of light was already considered by Einstein in the years
\citeyearNP{Einst:07}- \citeyearNP{Einst:11}.
Contrarily to first intuition, this does not contradict the special
theory of relativity (SR). `Constancy' of $c$, as far as SR is concerned,
refers to a limiting velocity with respect to Lorentz transformations
in one spacetime point; it does not mean $c$ has to be a constant function
in spacetime.
\citeN{Ran:04a} collected several citations
of Einstein that put into evidence that the principle of constancy (over spacetime)
of the speed of light is not a necessary consequence of the principle of 
relativity.\fo{Further comments are given in \citeN{Unz:07b}.}.

Considering variable atomic length and frequency standards $\lambda $
and $f$ \fo{Therefore $\lambda f =c$ is variable, too.} does nothing
else but influence our measurements of distances, i.e. the metric
of spacetime. A variable metric however manifests as
curvature.\fo{A
textbook example herefore 
 is  given  by \shortciteN{FeyII}, chap.~42.} It will be
discussed in sec.~\ref{art}, how the arising curvature can be brought in agreement
with the known tests of GR.

\subsection{Electromagnetic and Brans-Dicke-theories}
Shortly after GR was generally accepted, \citeN{Wil:21}
reconsidered Einstein's early attempts and deduced Newton's law
from a spacetime with a variable refractive index depending on the
gravitational field. This and related proposals like Rosen's
\citeyear{Ros:40} remained quite unknown until Dicke's paper
\citeyear{Dic:57} attracted much attention with the statement that
gravitation could be of electromagnetic origin. While the second
term in Dicke's index of refraction (eqn.~5 there) \be \e=
1+\frac{2 G M}{r c^2} \ee is related to the gravitational
potential of the sun, Dicke was the first to raise the speculation on
the first term having `its origin in the remainder of the matter
in the universe'. While this implementation of Mach's principle
was fascinating, in the following, more technical development of
the theory \cite{Bra:61}, a new parameter $\om$ was introduced
that
 later failed to agree with a reasonable experimental
value \cite{Rae:79}. Without Mach's principle, theories with a
variable refractive index, and therefore with a variable $c$,
were developed as `polarisable vacuum models' of GR. In particular,
\citeN{Put:99} showed a far-reaching agreement with all
experimental tests of GR known so far.

\subsection{Recent VSL theories}

Recently, Ra\~{n}ada (\citeyearNP{Ran:04a}) considered 
a VSL in the context of the Pioneer anomalous acceleration.
As an alternative to inflation models in cosmology,
variable speed of light (VSL) theories were developed by
Moffat (\citeyearNP{Mof:93}; \citeyearNP{Mof:02})
and Magueijo (\citeyearNP{Mag:00}; \citeyearNP{Mag:03}). These attempts had to face
some criticism that claimed that considering any variability of $c$
is just a trivial and obsolete transformation. Indeed, 
this criticism would be correct if spacetime were flat. A review
article by \citeN{Mag:03} that focusses on cosmological aspects gives
 a detailed answer to that wrong argument (sec.~2). Unfortunately,
 a lot of controversial discussion on VSL theories recently arose.
Some clarifying comments on this topic were given by
\citeN{Eli:03}. 
Since the present proposal refers to old established
ideas of Einstein, Dirac, Sciama and Dicke,
I shall not like to enter the actual discussion.

\subsection{Time and Length scales} 
The only reasonable way to define time and length scales is by
means of the frequency $f$ and wave length $\lambda $ of a certain atomic transition,
as it is done by the CODATA units. Thus by
definition, $c= f \lambda = 299792458 \frac{m}{s}$ has a fixed value
{\em with respect to the time and length scales\/} that locally however
may vary.\fo{Of course, such variation cannot be detected unless
we use spacetime curvature as indirect measurement.}
To analyze such situations in the following,  
we get 
from the definition, 
\be dc = \lambda df + f d \lambda ,  \ee 
\ni and 
\be
\frac{dc}{c}=\frac{df}{f} + \frac{d \lambda }{\lambda }, \label{cfl} 
\ee 
using
the product rule. Since we are interested in analyzing first-order
effects, we will use (\ref{cfl}) for finite differences $\delta$ as
well.

\subsection{Spatial variation of $c$.}
How can we reasonably speak of a variation of $c$~? Let's start
from eqn.~(\ref{cfl}). According to GR,  $\lambda $ is lower in the
gravitational field, {\em seen from outside\/}, i.e. at a large
distance from the field mass. Equally, frequency scales are lowered, 
and therefore, {\em time} scales raised. 
That means time runs slower, if we again observe from outside. 

According to (\ref{cfl}), we must assume $c$ to be lowered by
the double amount in the vicinity of masses, if we perform measurements
at large distances. This point of view is not that common because we usually do not
consider the value of $c$ at a distance, though we do  consider
$\lambda $ and $f$ (gravitational redshift!). For a detailed and elucidating discussion of this
viewpoint in the context of the radar echo delay, see \citeN{Wil}, p.~111 ff.

The lowered $c$ in the vicinity of masses is sometimes 
called `non-proper´ speed of light, whereas
the locally measured `proper' speed of light is a universal constant
\cite{Ran:04a}.


\section{Variable speed of light and tests of GR.} \label{art}

\subsection{Notation}
In order to fulfill (\ref{cfl}) while analyzing variations of
$c$, we are seeking a simple notation. 
In first order, all GR effects involve the relative change
$\frac{\delta x}{x}$ of the quantity $x$ of the amount 
$-\frac{G M}{rc^2}$. $\delta x = x_g-x_0$, where $x_0$ is the
  {\em locally measured\/}  quantity
and $x_g$ the quantity in the gravitational field, {\em
measured from the observer outside\/}. We shall abbreviate a
lowering in the gravitational field as $\da$, e.g. the well-known
slower running clocks with $\frac{\delta f}{f}=-\frac{G M}{r c^2}$ as
$f \da$, and equally the shortening of rods
$\frac{\delta \lambda }{\lambda }=-\frac{G M}{r c^2}$ as $\lambda \da$.
$\da \da$ stands for the double amount $-\frac{2G M}{r c^2}$, 
and $\ua$ indicates a relative increase, etc.\fo{Equivalently, $\da$
stands for a factor $1-\frac{G M}{r c^2}$. For factors close to 1,
$(1-\frac{G M}{r c^2})^2  \approx 1-\frac{2G M}{r c^2}$ holds, 
corresponding to $\da \da$.}
Eqn.~(\ref{cfl}) has to be fulfilled everywhere. No change is
indicated as $x \ra$. Only two hypotheses are needed for
describing the following experiments: 
\bi \i In the gravitational
field $c \da \da = f \da +\lambda \da$ holds. 
\i During propagation,
the frequency $f$ does not change. \ei

\subsection{The Hafele-Keating  experiment.} 
In \citeyearNP{HKe}, two atomic clocks were transported in aircrafts surrounding 
the earth eastwards and westwards in an experiment by Hafele and Keating.
 Besides the SR effect of moving clocks
that could be eliminated by the two flight directions, the results
showed an impressive confirmation of the first-order general
relativistic time delay. We briefly describe the result by $f
\da$, since the slower rate at which clocks run in a gravitational
field was measured directly. The accuracy was very much improved
by \citeN{Ves:80}. From this experiment alone it does not follow
yet $c \da \da f \da \lambda \da$, because $\lambda $ could even remain
unchanged. The following tests however leave no other possibility
but the above mentioned.

\subsection{Gravitational redshift and the Pound-Rebka experiment.}
Photons leaving the gravitational field of the earth show a slight
decrease in frequency at distances of about 20 m. This high-precision experiment
became possible after the recoil-free
emission of photons due to the M\"ossbauer effect
(\citeNP{Pou:60}; \citeNP{Pou:65}). Since
the experiment consists in absorption of photons $E= h f$, this is a true
$frequency$ measurement. Again, the frequency change $f \da$
was verified. For this point of view
it is important to remember the assumption that the
photon {\em does not\/} change its frequency
while propagating in space, but had a lowered rate $f$ already
at the emission.

The same effect, but measured as an optical wavelength shift, was
verified analyzing the emission spectra of the sun \cite{Sni:72}.
The situation at the emission is: 
\be c \da \da = f \da +\lambda \da
\ee 
During propagation $f=const.$ holds, and the local $c$ at
arrival remains per definition unchanged. Thus at the time of detection,
\be c \ra = f \da +\lambda \ua 
\ee 
holds, what indeed is observed. In
plain words, one can imagine the process as follows: consider a
photon travelling from its home star with huge gravitation to the
earth (with approximately zero gravity). While travelling through
different regions in space, it keeps its (lowered) frequency. The
photon does not keep in mind, so to speak, that its origin was a
cesium transition - in that case we would not note a redshift at
the photons arrival - but it maintains its frequency. Since this
happens at a space position outside the gravitational field where
$c$ is higher by the double amount (see (\ref{cfl})), the photon
has to adjust its $\lambda $, and raise it with respect to the value `at
home'. Since the adjustment to $c$ overcompensates the originally
lower $\lambda $, we detect the photon as gravitationally redshifted.

\subsection{Radar echo delay and light deflection.}

The above results, linked to eqn.~(\ref{cfl}), already suggest
a relative change of the speed of light of the amount
$c \da \da$ in the gravitational field. During propagation, $f$
again does not change.
The first idea of measuring this effect directly was launched
by  \citeN{Sha:64}.
Data of spectacular precision that agreed on the level of $10^{-3}$
with the theoretical prediction were collected by the Viking lander
missions \cite{Rae:79}. Recently, the Cassini mission delivered a
still better agreement \cite{WTB:04}.

Measuring the radar echo delay is only possible because the photon
maintains its frequency while travelling. Since $c$ is lowered by
the double amount in the vicinity of the star $c \da \da$, it has
to shorten its wave length accordingly, and while bypassing the
star \be c \da \da = \lambda \da \da f \ra \ee holds. There, the
photon's $\lambda $ is even shorter than $\lambda $ of a photon produced at
the star by the same atomic transition (see above, $\lambda \da$).
Naturally, the property which the photon has to leave at home for
sure, is `its' $c$, otherwise it wouldn't be considered as light
any more. As it is outlined by \citeN{Wil}, p.~111, too,
interpreting the radar echo delay as a local modification of $c$
is an uncommon, but correct interpretation of GR. All the tests
are in agreement with (\ref{cfl}).

A lower $c$ in the vicinity of masses creates light deflection
just as if one observes the bending of light rays towards
the thicker optical medium. Quantitatively, light deflection is
equivalent to the radar echo delay. A detailed calculation
for the total deflection $\Delta \ph$ yields
\be
\delta \ph = \frac{4 G M}{r c^2}.
\ee
Again the bending of light appears as curvature, therefore it is justified
to describe GR by a spatial variation of $c$, as it was already pointed
out by \citeNP{Einst:11}:
\bq
`From the proposition which has just been proved, that the velocity
of light in the gravitational field is a function of the place,
we may easily infer, by means of Huyghens's  principle,
that light-rays propagated across a gravitational field undergo deflexion'.
\eq

In the same article, Einstein obtained
only the  half value for the light deflection, because he considered
$c$ and $f$ as subject to variation, but not $\lambda $.

Independently from describing GR by a VSL theory,
in a given spacetime point there is only
one reasonable $c$, from whatever moving system one tries to measure.
This, but not more is the content of SR, which is not affected
by expressing GR by a variable $c$.

\subsection{Measuring masses and other quantities}

Not only time and length but every quantity is measured in units
we have to wonder about, once we allow changes of the measuring
rods. An interesting question is how to measure masses. Consider a
large field mass $M$ with gravitational field and at large
distance a test mass $m$. By Newton's 3rd law \be F_M = - F_m \ee
holds, thus the `force measured at a distance' is always of the
same amount as the `local force'. However, in \be a_M M = - a_m m
\label{new3} \ee $a_m$ is measured with the unchanged scales, and
$a_M$ with the scales inside the gravitational field. The
measurement of accelerations depends on time and length scales,
and the unit $\frac{m}{s^2}$ corresponds to a three times lower
value for the acceleration in gravitational fields, or $a \da \da
\da$ in the above notation. To remember, this holds if we observe
the behavior of an accelerated object at a large distance, i.e.
from outside of the gravitational field. Consequently, the
measurement of masses is influenced as well. Due to (\ref{new3}),
masses are proportional to reciprocal accelerations and $ m \ua
\ua \ua$ holds.

Similar considerations lead to $v \da \da$ for velocities (as $c$), and
$l \ra$ for the angular momentum. This last observation will turn out useful for keeping
Planck's constant $h$ spatially constant, since it has units $kg \frac{m^2}{s}$. 
\fo{In agreement, \citeN{Dic:57}, p.366, and \citeN{Put:99}, eqns.~8-13, with
different arguments.}

Table~I gives an overview on the change of quantities. One $\da$
abbreviates a relative decrease of the quantity $\frac{\delta x}{x} = -\frac{G M}{r c^2}$
in the gravitational field, or a factor $(1-\frac{G M}{r c^2})$.

\vspace{0.5cm}

\begin{tabular}{|lcccr|}
\hline
Quantity & symbol &unit & rel. change &\\
\hline
Frequency  & f  & $\frac{1}{s}$ & $\da$ &\\
Time  & t  & $s$ & $\ua$ &\\
Length  & $\lambda $  & $m$ & $\da$ &\\
Velocity  & v  & $\frac{m}{s}$ & $\da \da$ &\\
Acceleration  & a  & $\frac{m}{s^2}$ & $\da \da \da$ &\\
Mass  & m  & $kg$ & $\ua \ua \ua$ &\\
Force  & F  & $N$ & $\ra$ &\\
Pot. energy  & $E_p$  & $Nm$ & $\da$ &\\
Ang. mom.  & l  & $\frac{kg m^2}{s}$ & $\ra$ &\\
\hline
\end{tabular}\\

Table~I. Overview on the change of quantities.

\vspace{0.1cm}

\subsection{Advance of the perihelion of \\ Mercury}
In the Kepler problem, the Lagrangian is given by
\be
L = \frac{m}{2} (\dot r^2 + r^2 \dot \ph^2) + \frac{G M m}{r},
\ee
which after introducing the angular momentum ${l = m r^2 \dot \ph}$ transforms to
\be
L = \frac{1}{2} m \dot r^2 + \frac{l^2}{2 m r^2} + \frac{G M m}{r}. \label{Kep}
\ee
Now we have to consider the relative change of the quantities as given above. $m \ua \ua \ua$
however refers to a test particle at rest. The kinetic energy causes an additional,
special relativistic mass increase $\ua$ (see sec.~5.1 and 5.2 below in detail), thus
in eq.(\ref{Kep}), $m \ua \ua \ua \ua$ holds, while $r \da$ and $l \ra$. Hence, the middle term
changes as $\da \da$ in total, which means that $m$ is effectively multiplied by a factor
$(1-\frac{2GM}{r c^2})$.\fo{As one easily checks, the first and the last terms instead do not need 
a correction.}
The arising  $-\frac{GM l^2}{m r^3 c^2}$ represents the well-known correction which leads to the secular
shift 
\be
\Delta \phi = \frac{6 \pi G M}{A (1-\e^2) c^2},
\ee
$A$ being the semimajor axis and $\e$ the eccentricity of the orbit.

\section{Newton's law from a variable $c$.}

Since time and length measurement effects of GR can be described
by a spatial variation of $c$, one is tempted to try a description
of all gravitational phenomena in the same framework.
However, $c \da \da$ (see sect.~4.1) requires
\be
\delta (c^2)= 2 c \delta c = -\frac{4 G M}{r}. \label{4pot}
\ee
This differs by a factor 4 from Sciama's
original idea  and suggests a Newtonian gravitational potential of the
form
\be
\ph_{Newton} = \frac{1}{4} c^2.
\ee

\subsection{Kinetic energy.}
As it is well-known from special relativity (SR), the mass increase due to
relativistic velocities is, applying Taylor series to the square root,
\bea
\delta m = m(v) -m_0 = \\ \nonumber
m_0 \bigl(\frac{m_0}{\sqrt{1-\frac{v^2}{c^2}}}-1 \bigr)
\approx m_0 (\frac{1}{2} \frac{v^2}{c^2}). \label{ekinSR}
\eea
If we apply conventional energy conservation $E_k= E_p$ and start at
$r= \infty$, $v=0$,  with
$\frac{1}{2} v^2= \frac{G M}{r}$ follows
\be
\frac{\delta m}{m} = \frac{G M}{r c^2}, \label{Ekin}
\ee
for the relative mass increase due to kinetic energy, corresponding to $\ua$.

\subsection{Energy conservation.}
A quite interesting consequence of the foregoing calculation arises
if we consider the quantity $E= m c^2$ of a test particle moving
into a gravitational field. Recall that time and length measurements
influence all quantities, thus $\frac{\delta c^2}{c^2} = -\frac{4 G M}{r c^2}$
or $c^2 \da \da \da \da$. This is partly compensated by $m \ua \ua \ua$
as outlined above. There is one relative increase of the amount
 $\frac{G M}{r c^2}$ left - the part in eqn.~(\ref{Ekin}) due to
the increase in kinetic energy!

This means, the total energy $E= m c^2$ of a test particle appears as a
conserved quantity during the motion in a gravitational field.
In first order, we may write
\be
\delta(m c^2) = c^2 \delta m + m \delta (c^2) =0. \label{PR}
\ee
Roughly speaking, the left term describes the increase in
kinetic energy due to the relativistic $\delta m$ and the right term
the decrease of potential energy in a region
with smaller $c^2$.
To be quantitative, however, the change in $m$ contributes to
the change in potential energy such that $E_p \da$.
The change in of the right term,
$m \delta(c^2)$ amounts to $-\frac{4 G M}{r c^2}$, four times as
much as needed to compensate $d E_k$.
Three parts of it are compensated by an increase of
$m, \ua \ua \ua$ due to mass scales,
the remaining $\frac{1}{4} m d (c^2)$ is converted into
kinetic energy:
\be
m c^2 \ra m (c^2 \da \da \da \da) + c^2 (m \ua \ua \ua) + E_k
= E_p + E_k
\ee
In summary, there is nothing like a gravitational energy, just
$E=m c^2$. Kinetic energy is related to a change of $m$, and potential energy to
a change of $c^2$.
As it should be, photons do not behave differently, since they conserve their
energy $h f$ as well during propagation.

\subsection{Foundation of the equivalence principle.}

While GR provides most elegantly a formalism that incorporates the
equality of inertial and gravitating masses, the question as to the
deeper reason of this outstanding property of matter ('I was utterly
astonished about its validity', \citeNP{EinstWB}) is still open.
Why does the elementary property of matter, inertial resistance to accelerations, 
act at the same time as a `charge' of a particular interaction~?
Or, equivalently: why does kinetic energy show the same proportionality
to $m$ as potential energy in a gravitational field~?

If one supposes that the total energy of a particle can always
be written as $E=m c^2$, the differentiation in eqn.~(\ref{PR})
shows that both terms have to be proportional to the test mass $m$.

One fourth of the first term describes the (relativistic) increase of kinetic
 energy,\fo{$\frac{3}{4}$ of it are due to the mass increase $m \ua \ua \ua$. }
which is usually approximated as $\frac{1}{2}m v^2$. The remainder
plus the right term is the change in potential energy.
 Since $dm$ is proportional to $m$,
{\em both\/} terms are obviously proportional to the mass $m$ of the test
particle. There is no reason any more to wonder why inertial and gravitating mass are
of the same nature. Describing gravity with a varying speed of light
leads to the equivalence principle as a necessary consequence. The deeper
reason for this is that gravitation is strictly speaking not an interaction
between particles but just a reaction on a changing $c$.

\section{Dependence of $c$ on the mass distribution in the universe} \label{sciama2}


\subsection{Test masses and field masses.}

While dealing with the equivalence principle, there is a subtle difference
between test and field masses. The above derivation holds for the former only;
all spectacular tests of the EP, starting from E\"otv\"os to recently proposed
satellite missions, deal with test masses.

An entirely new question is how matter does influence $c$ in its
vicinity. Taking the proposal $\ph =\frac{1}{4} c^2$ seriously,
there are no more masses creating fields, because once you give up
Newton's $F= \frac{GMm}{r^2}$, the force does not need to depend
on a product of masses. The modification of $c$ could depend
instead on quantities which are approximately correlated to mass
like the baryon number. E\"otv\"os-like experiments could not
detect that, but there could be material dependencies in torsion
balance experiments.

\subsection{General considerations on the functional dependence of $c$.}

We are therefore seeking a Machian formula which describes
explicitly the dependence of $c$. The most general form
would be $c(m_i, \vec r_i, \vec v_i, \vec a_i, t)$, $i$ indicating a single
mass point in the universe.
Here I do not give detailed reasons why I exclude anisotropy
and an explicit dependence on $v_i, a_i$ and $t$.\fo{This would be unusual,
but not a priori senseless. \citeN{Bar:02} recently gave an interesting proposal
for $G$ depending on $v_i$.} The resulting hypothesis
\be
c^2= c_0^2 \ f(m_i, r_i) \label{chyp},
\ee
$c_0$ being the speed of light in some preferred condition of the universe,
has however to face the problem how to make the argument dimensionless.
Again, the only acceptable hypothesis seems to measure $m_i$ and $r_i$
either in multiples of $m_p, r_p$ of elementary particles (here protons) or as
fraction of the respective quantities of the universe $m_u, r_u$.
In both (or considering combinations, even four) cases, simple algebraic 
relations of $m_i$ and $r_i$ yield 
a very large or very small number.
For reasons that will become clear later, I prefer the choice $m_p, r_p$.
To avoid that $f(m_i, r_i)$ in  (\ref{chyp}) is of the order
$10^{40}$, one has to introduce a function like $\ln$.

\subsection{Proposal for the explicit dependence of $c$.}
Obviously, the $m_i$ should enter in an additive way, and the influence
of each mass $i$ should decay with $r_i$.
For these reasons, $\ln \sum \frac{m_i}{r_i}$ is likely to be a component of
eqn.~(\ref{chyp}), as one could suspect immediately from Sciama's
original proposal (\ref{sum}). Taking into account that $c$ has to decrease
in the vicinity of masses, one of the most simple possibilities\fo{The
alternative $c^2= - c_0^2 \ln \S$, while yielding similar results,
appears less natural since it requires $m_u$ instead of $m_p$.} is the formula
\be
c^2(m_i, r_i) = \frac{c_0^2}{\ln \sum \frac{m_i r_p}{r_i m_p}} =:
\frac{c_0^2}{\ln \S}. \label{sci2}
\ee

\subsection{Recovery of Newton's law}
For the acceleration of a test mass
\be
|a|= |\frac{1}{4} \nabla c^2| = \frac{c_0^2}{4 (\ln \S)^2}
\frac{\sum \frac{m_i}{r_i^2}}{\sum \frac{m_i}{r_i}}
\ee
holds, yielding the inverse-square law.
 The gravitational `constant' can be expressed as
\be
G= \frac{c_0^2}{4 (\ln \S)^2} \frac{1}{\sum \frac{m_i}{r_i}}=
\frac{c^2}{4 \ln \S} \frac{1}{\sum \frac{m_i}{r_i}}.
\label{Gsc2}
\ee

which differs by a numerical factor $4 \ln \S$ from Sciama's proposal.
It should be noted that the term $\sum \frac{m_i}{r_i}$ contained in $G$
is approximately constant. The contributions from earth, sun, and milky way, 
$9.4 \cdot 10^{17}$,  $1.33 \cdot 10^{19}$, and $7.5 \cdot 10^{20}$ (in $kg/m$),
are minute compared to the distant masses in the universe that 
approximately\fo{Taking into account that all matter could be visible (sec.~7.1),
the universe term is still dominant ($5.4 \cdot 10^{24}$).} amount 
to $1.3 \cdot 10^{27}$. This was first 
observed by \citeN{Sci:53}, p.~39.
Thus slight variations due to motions in the solar system are far below the 
accuracy of current absolute $G$ measurements ($\Delta G/G =1.5 \cdot 10^{-4}$).

\subsection{Gravitational potential.}

The gravitational $\ph$ potential of classical mechanics
arises in first order from eqn.~(\ref{sci2}). Taking the example $\ph_s=
\frac{G M_{\circ}}{R_{\circ}}$ of the sun, and
let  $\S^{'}$ be the  sum as defined in (\ref{sci2}) without the sun,
then, $\Delta$ denoting a difference,
\bea
\Delta c^2 = c_0^2 (\frac{1}{\ln \S^{'}}-\frac{1}{\ln \S}) = \\ \nonumber
\frac{c_0^2}{\ln \S \ \ln \S^{'}} (\ln \S -\ln \S^{'}) =
\frac{c^2}{\ln \S^{'}} \ln \frac{\S}{\S^{'}} = \\ \nonumber
\frac{c^2}{\ln \S^{'}} \ln (1+\a),
\eea
holds, whereby $\a$ is defined as  $\frac{M_{\circ}}{R_{\circ} \sum \frac{m_i}{r_i}}$.
The Taylor series of the $\ln$, after applying (\ref{Gsc2}),
 yields in first approximation
\be
\Delta c^2 = \frac{c^2}{\ln \S^{'}} \a =
\frac{4 G M_{\circ}}{R_{\circ}} = 2 \ c \Delta c =: 4 \ph_1,
\ee
$\ph_1$ being the first-order approximation of the gravitational potential of
the sun.

Though the present proposal differs from Sciama's by a numerical
factor, the comment on $G$ given on page 39 in the 1953 paper
still applies: \bq `... then, local phenomena are strongly coupled
to the universe as a whole, but owing to the small effect of local
irregularities this coupling is practically constant over the
distances and times available to observation. Because of this
constancy, local phenomena appear to be isolated from the rest of
the universe.' \eq

\subsection{Taking away the mystery from Mach}

There is another satisfactory aspect of a gravitational potential
of the form $\frac{1}{4} c^2$. A Machian dependence of Newton's
constant $G$ which is determined by the distribution of all masses
in the universe is in a certain sense beautiful, but after all
somewhat mysterious. How does a free falling apple feel the mass
distribution of the universe~? According to Sciama's
interpretation, its acceleration is much slower because there are
so many galaxies besides the milky way.

On the other hand, assumed that the gravitational potential is
just $\frac{1}{4} c^2$, it has been our belief in the universality
of Newton's law and the somewhat naive 
extrapolation of the
potential to $\sum_{i}  \frac{G m_i}{r_i}$ that created so much surprise
while rediscovering eqn.~(\ref{sum}).

Despite the fact that the potential $\sum_{i} \frac{G m_i}{r_i}$ is
a purely mathematical tool which cannot be observed directly, we have
internalized it so much that $G$ {\em seems\/} to depend on the 
mass distribution. In fact, if gravity depends just on $c$ instead, $G$
turns out as kind of an artifact.

\section{Discussion of consequences and observational facts}

In the meantime, the observational status of cosmology has improved dramatically.
Besides the unresolved problems sketched briefly in sec.~1.2, the increasing
precision in measuring the mass of the universe has revealed further riddles.

\subsection{Visible matter and flatness}

The approximate coincidence $\frac{c^2}{G} \approx \frac{m_u}{r_u}$ is
usually expressed as critical density of the universe
\be
\r_c = \frac{3 H_0^2}{8 \pi G}. \label{FriLem}
\ee
A universe with $\r_u >\r_c$ is called closed, and open in the case $\r_c >\r_u$.
An analysis of the evolution of the universe tells that if $\Om := \frac{\r_u}{\r_c}$
is about the order of 1 at present,  it must have been
as close as $10^{-60}$ to 1 at primordial times. For that, many cosmologists
believe $\Om$ has to be precisely 1, but a cogent theoretical reason is still
missing.
Moreover, observations clearly indicate that the fraction of visible matter,
$\Om_v$ is about 0.01, with a large uncertainty.

From (\ref{Gsc2}) and the assumption of an homogeneous universe,
elementary integration over a spherical volume yields
$\sum \frac{m_i}{r_i} \approx \frac{3 m_u}{2 r_u}$, and therefore
\be
m_u \approx \frac{c^2 r_u}{6 G \ln \S}
\ee
holds.
If one assumes $r_p =1.2 \cdot 10^{-15} m$, $m_p =1.67 \cdot 10^{-27} kg$
(the proton values), and taking the recent measurement
of $H_0^{-1} = 13.7 \ Gyr$ that leads to $r_u =1.3 \cdot 10^{26} m$,
$m_u$  is in approximate agreement with the observations,
and corresponds to the fraction of visible matter $\Om_v = 0.004$.
Moreover,  there is no more reason to wonder about the approximate
coincidence (\ref{FriLem}), since it is a consequence
of the variable speed of light (\ref{sci2}), if one uses $\r=\frac{3 m_u}{4 \pi r_u^3}$
and $r_u H_0 \approx c$.

Newtonian gravity deduced from a variable speed of light can thus
give an alternative explanation for the flatness problem and may
not need dark forms of matter (DM, DE) to interprete cosmological
data.

\subsection{The horizon problem}
If we consider the sum in (\ref{sci2}) for cosmological time scales,
we know very little about its time evolution. In particular, we know
about $r_i(t)$ only the momentary 
Hubble expansion $\dot r_i(13.7 Gyr)$.

It is tempting however to speculate about an increasing number of
$m_i$ dropping into the horizon, thus lowering $c$ during the
cosmological evolution. In this picture, the big bang could happen
at a horizon of the size of an elementary particle ($\S =1$) with
infinite $c$, the logarithmic dependence causing however a
very rapid decrease to the almost constant actual value. A
decreasing speed of light was recently considered by \citeN{Mag:00}
as an alternative scenario to inflation.

\subsection{Time evolution of $G$.}
The increase of the number of $m_i$ is also necessary for
compensating the increase of the $r_i$. It is at the moment not
excluded that $G$ in (\ref{Gsc2}) could meet the quite restrictive
observational evidence for $\dot G \approx 0$ \cite{Uza}.

\subsection{Temporal evolution of $c$.}
In sec.~4, a spatial variation of $c$ was considered describing
effects that occurred while $\dot c \approx 0$. As it can be
deduced from Maxwell's equations\fo{Of course, I do not
distinguish between an electromagnetic $c_{EM}$ and a spacetime
$c_{ST}$, see \citeN{Eli:03}.}, in this case a photon keeps its
frequency $f$ unchanged. On the other hand, if we assume $\dot c
<0$ during cosmological evolution in an almost isotropic universe
with vanishing gradients of $c$, Maxwell's equations require a
constant $\lambda $ during wave propagation, in the foregoing notation
$c \da \da = f \da \da \lambda \ra$. Since a photon keeps its original
$\lambda $, the wavelength appears longer with respect to the
 measuring rods that evolve according to $c \da \da = f \da \lambda \da$.
\citeN{Dic:57}, p.~374, suspected this result to be related to
the cosmological redshift.

\section{Outlook}

Galileo's $F=m g$, a formula that today could enter the elementary
schools, has been a quite remarkable discovery, though describing
effects on earth only. Newton's generalizing of this formula and
describing celestial mechanics of the solar system was even more
difficult.

I fear that the hierarchy of structures earth - solar system -
galaxy - cosmos may require a corresponding hierarchy of
physical laws that cannot be substituted by introducing new
parameters, data fitting and numerical simulations.

Due to satellite technique, improved telescopes and the computer-induced
revolution in image processing we are collecting
data of fantastic precision.
At the moment it is unlikely that our theoretical understanding
on the galactic or cosmologic scale can match the amount of data. 

Though the range of the universe we know about since the discovery
of GR has increased dramatically, we usually extrapolate
conventional theories of gravity to these scales. Physicists have
learned a lot from quantum mechanics, but the historical aspect of
the lesson was that the 19th-century extrapolation of classical
mechanics over 10 orders of magnitude to the atomic level was a
quite childish attempt.

Under this aspect, the amount of research in cosmology which is done nowadays
on the base of an untested extrapolation over 14 orders of magnitude
is a quite remarkable phenomenon.

The wonderful observational data of the present should not lead us
to neglect the theoretical efforts of scientists in the past.
Sometimes one can learn more from the `blunders' of deep thinkers
like Einstein,
Lord Kelvin, Mach or Sciama than from actual theories in fashion. 

Since it deals with the very basics, it may be to early to
try to test the present proposal with sophisticated data.
It is merely an attempt to open new roads
than a complete solution to the huge number of observational riddles.\fo{An
excellent overview on the observations modified gravity has to match
gives \shortciteN{Agu:01a}.}

Thus I consider sec.~7 as minor and preliminary results; besides the
reproduction of GR and Newton, the more
satisfactory aspect of this proposal seems the motivation for
the equivalence principle, the implementation
of Mach's principle and the `removal' of the constant $G$.

\paragraph{Acknowledgement.} 
I am deeply indebted to Karl Fabian for inspiring discussions and
for correcting many erroneous ideas of mine. The intelligent
proofreading of Hannes Hoff substantially clarified many things.
Comments of an unknown referee were helpful for v3.

\end{document}